\documentclass[aps,pra,showpacs,twocolumn, superscriptaddress, longbibliography]{revtex4-1}

\usepackage[utf8]{inputenc}
\usepackage{amsmath}
\usepackage{booktabs}
\usepackage{siunitx}
\usepackage{float} 
\usepackage{graphicx}
\usepackage{dcolumn}
\usepackage{bm}
\usepackage[colorlinks]{hyperref}
\usepackage{chemformula}
\usepackage{graphicx}
\usepackage{amsmath,amssymb}  
\usepackage[T1]{fontenc}      
\usepackage{lmodern}          

\newcommand{\rev}[1]{{\color{black}#1}}

\newcommand{\jjred}[1]{{\color{black}#1}}

\begin{document}

\title{Shaping the spectra of second harmonic laser by spatial focusing}

\author{Yan-Ming Liu}
\affiliation{Hubei Key Laboratory of Optical Information and Pattern Recognition, Wuhan Institute of Technology, Wuhan 430205, China}
\author{Tao Xie}
\affiliation{Key Laboratory of Low-Dimensional Quantum Structures and Quantum Control of Ministry of Education, Department of Physics and Synergetic Innovation Center for Quantum Effects and Applications, Hunan Normal University, Changsha 410081, China}
\author{Xiangying  Hao}
\affiliation{Hubei Key Laboratory of Optical Information and Pattern Recognition, Wuhan Institute of Technology, Wuhan 430205, China}
\author{Rui-Bo Jin}
\email{Corresponding author: jrbqyj@foxmail.com}
\affiliation{Hubei Key Laboratory of Optical Information and Pattern Recognition, Wuhan Institute of Technology, Wuhan 430205, China}
\affiliation{Key Laboratory of Low-Dimensional Quantum Structures and Quantum Control of Ministry of Education, Department of Physics and Synergetic Innovation Center for Quantum Effects and Applications, Hunan Normal University, Changsha 410081, China}

\date{\today}

\begin{abstract}

\rev{Accurately controlling the spectral width of second harmonic generation (SHG) is critical in applications such as nonlinear microscopy, ultrafast spectroscopy, and quantum optics, where tailored spectral properties are required. Although it is known that the focal length of the lens used to focus the pump beam influences the SHG spectrum, a simple and precise theoretical understanding of this effect has been lacking. Here, we develop a theoretical model to describe how pump focusing conditions affect SHG spectral broadening, and experimentally validate the model using BBO crystals with lenses of 50 mm and 200 mm focal lengths. Our results establish a practical method for engineering the SHG spectral width through pump focusing, providing valuable guidance for designing nonlinear optical systems with specific spectral requirements.}

\end{abstract}

\maketitle


\section{Introduction}
Second-harmonic generation (SHG) is a fundamental nonlinear optical process wherein two photons of identical frequency interact within a nonlinear material, generating a new photon at twice the frequency. As one of the most extensively studied nonlinear phenomena, SHG plays a pivotal role across a broad range of disciplines \cite{denev2011probing}. In fundamental scientific research, SHG is essential for ultrafast spectroscopy and quantum optics \cite{orenstein2012ultrafast, Xiao2026Ultrabroadband,He2025Coherent,scully1997quantum}. In technological applications, it enables frequency conversion for laser sources and supports optical communications \cite{petrov2015frequency, yariv2007photonics}. In the life sciences, SHG facilitates advanced biological imaging and microscopy \cite{jin2011application, cicchi2013molecular}, while in materials science, it is widely used for material characterization  \cite{busch2006photonic}. 
The remarkable capability of SHG to generate coherent light at new wavelengths has thus established it as an indispensable tool in both fundamental research and practical applications \cite{Niu2023COL, Niu2021JOSAB}.

Several parameters are commonly used to characterize the performance of SHG, including output power, conversion efficiency, temporal width, spectral width (bandwidth), beam quality, polarization state, wavelength tunability \cite{chaitanya2009high, Li2014LowPower, Han2017Efficient, Zhang2026Vacuum}. Among these, the spectral width is a particularly critical parameter and it is essential to  precise characterize and control it. The spectral properties of SHG have therefore attracted considerable attention. 
\rev{
The study of SHG began with the pioneering observation in quartz by Franken \textit{et al.} in 1961 \cite{franken1961} (without phase matching), followed by the first phase-matched SHG in KDP crystals by Giordmaine in 1962 \cite{giordmaine1962}, and the development of foundational quasi-phase-matching theories by Armstrong  \textit{et al.}  in the same year \cite{armstrong1962}. The effect of focusing geometry on SHG was first systematically investigated by Bjorkholm in 1966, who theoretically and experimentally studied how focusing influences SHG efficiency in ADP crystals \cite{Bjorkholm1966PR}.
\jjred{In 1968, Boyd and Kleinman developed a more comprehensive theoretical framework for optimizing SHG with focused Gaussian beams \cite{boyd1968parametric}}.
\jjred{Building on these foundational studies}, Blit \textit{et al.} (1979) experimentally explored the angle bandwidth of SHG using focused beams in ADP crystals \cite{blit1979wavelength}. However, the theoretical models developed in these studies were based on monochromatic lasers.
With the advancement of femtosecond laser technology, attention has shifted towards SHG using broadband lasers. Recent experiments by Chaitanya \textit{et al.} (2014) demonstrated that tighter focusing of femtosecond laser pulses in thick BIBO crystals leads to broader SHG spectra, while looser focusing results in spectral narrowing \cite{Chaitanya2014OL}. Despite these experimental observations, a simple and precise theoretical model that quantitatively describes the dependence of SHG spectral width on pump focusing conditions remains unavailable.
}

In this work, we develop a theoretical model to calculate the spectral distribution of SHG, taking into account the effects of lens focal length, pump bandwidth, and beam radius. To validate the model, we experimentally measured the SHG spectra from a BBO crystal using lenses with focal lengths of 50 mm and 200 mm. The experimental results show excellent agreement with the theoretical predictions. Our approach offers a simple and effective method for controlling the spectral width of the second-harmonic output by adjusting the focusing conditions.

\begin{figure*}[!thp]
    \centering
\includegraphics[width= 1.0\textwidth]{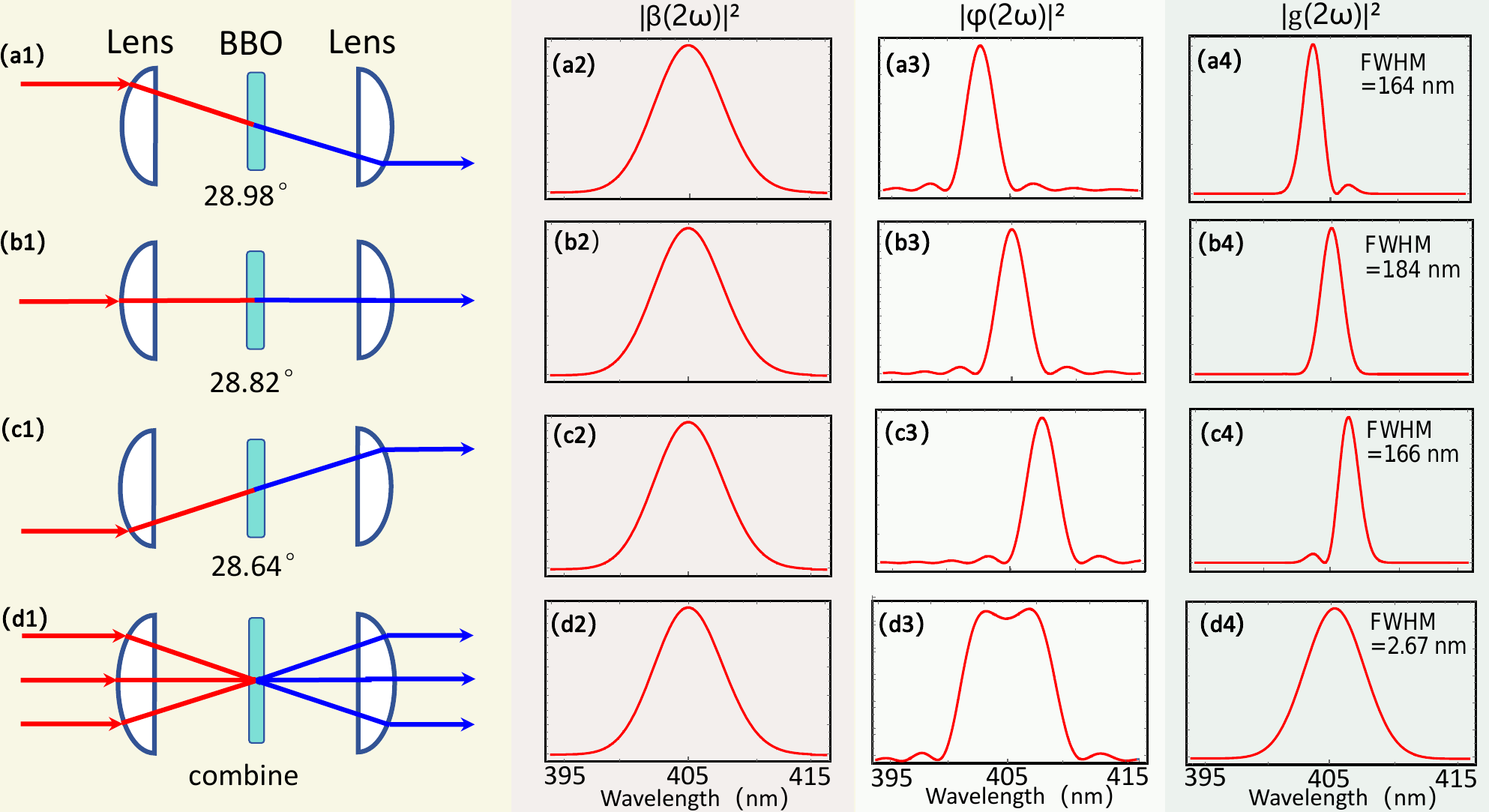}
\caption{
Conceptual illustration of this research. (a1–d1) The first column shows the SHG setup, where the fundamental laser is focused onto the nonlinear crystal using a lens at different incident angles. 
(a2–d2) The second column presents the intensity of the pump envelope function $\beta(2\omega)$; all distributions are identical  since the same pump laser is used in each case.
(a3–d3) The third column shows the intensity of the phase-matching function $\varphi(2\omega)$, for each angle.
(a4–c4) The fourth column displays the intensity of the SHG spectral distribution $|g(2\omega)|^2= |\beta(2\omega)|^4 \times |\varphi(2\omega)|^2$ for each angle.
(d3)  is the modulus squared of the combination of the amplitudes of (a3), (b3), and (c3).
Compared to (a4–c4), (d4) exhibits a broader spectral distribution.
}
\label{fig::concept}
\end{figure*}

\section{Theory}
SHG process is described by the theory of three-wave mixing. Under the typical assumptions of plane-wave propagation, a slowly varying envelope approximation, and negligible optical losses, the intensity of the generated second-harmonic field can be expressed as \cite{aghigh2023}:
\begin{equation}\label{Eq:1}
I_{2\omega} = \frac{8\omega^2 d_{\text{eff}}^2 L^2}{n_{2\omega} n_{\omega}^2 c^3 \varepsilon_0} \operatorname{sinc}^2\! \left( \frac{\Delta k L}{2} \right) I_{\omega}^2,
\end{equation}
where $I_{2\omega}$ and $I_{\omega}$ are the intensities of the second-harmonic and fundamental beams, respectively; $d_{\mathrm{eff}}$ is the effective nonlinear coefficient of the crystal; $n_{2\omega}$ and $n_{\omega}$ denote the refractive indices at the second-harmonic and fundamental frequencies; $c$ is the speed of light in vacuum; $\varepsilon_0$ is the vacuum permittivity; $\Delta k = k_{2\omega} - 2k_{\omega}$ is the phase-mismatch parameter; and $L$ is the interaction length in the crystal.

To describe the spectral profile of the frequency-doubled light, we consider the amplitude distribution $g(2\omega)$. It can be factored into the product of the pump spectral envelope $\beta(2\omega)$ and the phase-matching function $\phi(2\omega)$~\cite{SYCS202412008}:
\begin{equation}
g(2\omega) = \beta^2(2\omega) \times\phi(2\omega).
\end{equation}
For a transform-limited fundamental pulse, the envelope $\beta(2\omega)$ is typically modeled as a Gaussian:
\begin{equation}
\beta(2\omega) = \exp\left[ -\frac{1}{2} \left( \frac{\omega - \omega_0}{\sigma_0} \right)^2 \right],
\end{equation}
where $\omega_0$ is the central frequency and $\sigma_0$ the spectral width of the fundamental beam. 
The phase-matching function is:
\begin{equation}
\phi(2\omega) = \operatorname{sinc}\!\left( \frac{\Delta k L}{2} \right)\operatorname{exp}\!\left( \frac{ i \Delta k L}{2} \right).
\end{equation}
Consequently, the second-harmonic intensity in Eq.~(\ref{Eq:1}) can be recast in terms of these spectral functions as
\begin{equation}
I_{2\omega} = c_0 \, |\beta(2\omega)|^4  \times   |\phi(2\omega)|^2,
\end{equation}
with the prefactor $c_0 = \frac{8\omega^2 d_{\mathrm{eff}}^2 L^2}{n_{2\omega} n_{\omega}^2 c^3 \varepsilon_0}$. For a birefringent crystal cut at an angle $\theta$, the phase mismatch $\Delta k$ depends on the wave vectors of the fundamental ($k_{\omega}$) and second-harmonic ($k_{2\omega}$) waves:
\begin{equation}
\Delta k = k_{2\omega}(\theta) - 2k_{\omega}(\theta).
\end{equation}
Crucially, as illustrated in Fig.~\ref{fig::concept}, the phase-matching function $\phi(2\omega)$ becomes angularly dependent. When the pump beam is focused, a distribution of incident angles contributes to the overall signal. The spectral superposition of these different angle components results in a broadening of the second-harmonic spectrum.

\begin{figure*}[!thp]
\centering
\includegraphics[width= 1\textwidth]{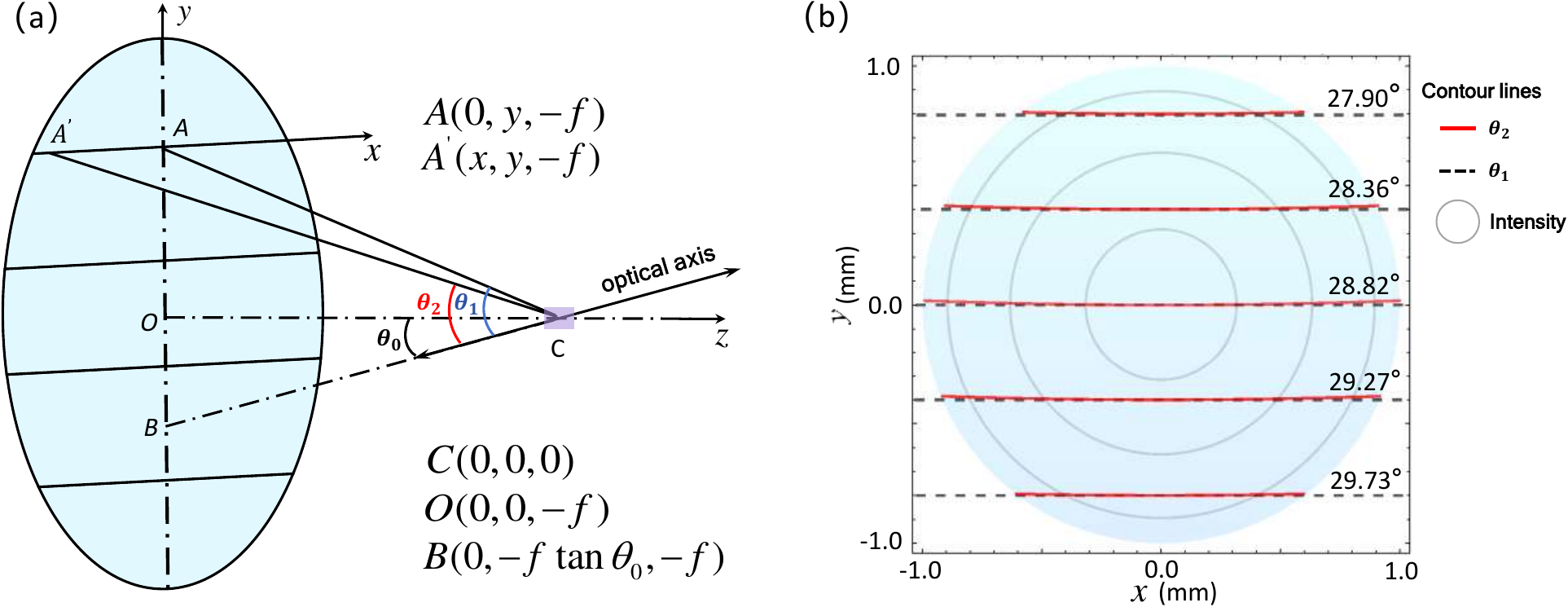}
\caption{Model of SHG under tight focusing conditions. (a) All rays from the cross section of the pump beam are focused onto the nonlinear crystal at point C. (b) Calculated distribution of incident angles across the cross section of the pump beam.
}
\label{fig::model}
\end{figure*}

We now derive the distribution of incident angles across the transverse cross-section of the pump beam. The coordinate system is defined in Fig.~\ref{fig::model}(a). The nonlinear crystal is positioned at point \(C(0, 0, 0)\), and the center of the focusing lens, which is also the center of the beam cross-section, is located at point \(O(0, 0, -f)\). Let \(A'(x, y, -f)\) denote an arbitrary point on the input plane of the lens, and consider a reference point \(A(0, y, -f)\) on the \(y\)-axis for comparison. The optical axis is defined by the line \(\overrightarrow{CB}\). The incident rays from points \(A\) and \(A'\) to the crystal are therefore given by vectors \(\overrightarrow{CA}\) and \(\overrightarrow{CA'}\), respectively. These vectors are expressed as:
\begin{align*}
    \overrightarrow{CB} &= \begin{pmatrix} 0, & -f\tan\theta_0, & -f \end{pmatrix}, \\
    \overrightarrow{CA} &= \begin{pmatrix} 0, & y, & -f \end{pmatrix}, \\
    \overrightarrow{CA'} &= \begin{pmatrix} x, & y, & -f \end{pmatrix},
\end{align*}
where \(\theta_0\) is the crystal cut angle (phase-matching angle for the on-axis ray).

The phase-matching angles \(\theta_1\) and \(\theta_2\) for rays originating from points \(A\) and \(A'\), respectively, are defined as the angles between each ray and the optical axis (\(\overrightarrow{CB}\)). They are calculated using the dot product formula:
\begin{align}
    \theta_1 &= \angle ACB
    = \arccos\left( \frac{\overrightarrow{CA} \cdot \overrightarrow{CB}}{|\overrightarrow{CA}| \, |\overrightarrow{CB}|} \right) \notag \\
    &= \arccos\left( \frac{-f y \tan\theta_0 + f^{2}}{f\sqrt{\tan^{2}\theta_0 + 1} \, \sqrt{y^{2} + f^{2}}} \right),
    \label{eq:theta1}
\end{align}
\begin{align}
    \theta_2 &= \angle A'CB
    = \arccos\left( \frac{\overrightarrow{CA'} \cdot \overrightarrow{CB}}{|\overrightarrow{CA'}| \, |\overrightarrow{CB}|} \right) \notag \\
    &= \arccos\left( \frac{-f y \tan\theta_0 + f^{2}}{f\sqrt{\tan^{2}\theta_0 + 1} \, \sqrt{x^{2} + y^{2} + f^{2}}} \right).
    \label{eq:theta2}
\end{align}

With these formulas, we can calculate the distribution of $\theta_1$ and $\theta_2$.
Figure~\ref{fig::model}(b) shows the calculated results. In this calculation, we consider a SHG process in a type-I phase-matched BBO from 810 nm to 405 nm, with $\theta_0$=28.82°.
We also assume a lens focal length of f = 50 mm, and a cross-section range   $-1 \leq x \leq 1$ and $-1 \leq y \leq 1$. 
It is interesting to find that the contour plot of $\theta_2$   almost overlaps with that of $\theta_1$, both distributed along horizontal lines. 
This phenomenon can be intuitively  explained as follows: since the beam size is much smaller than the focal length, the $x$ value has almost no effect on $\theta_2$ in Eq.~(8). 
As a result, all points along the same horizontal line share the same phase-matching angle; in other words, $\theta_2$ can be replaced by $\theta_1$.
By utilizing this feature, the calculation can be  significantly simplified, since the incident angle distribution at any point $A'$ can be replaced by that at point  $A$.

\section{Numerical Simulation}

\begin{figure*}[]
\centering
\includegraphics[width= 1.0\textwidth]{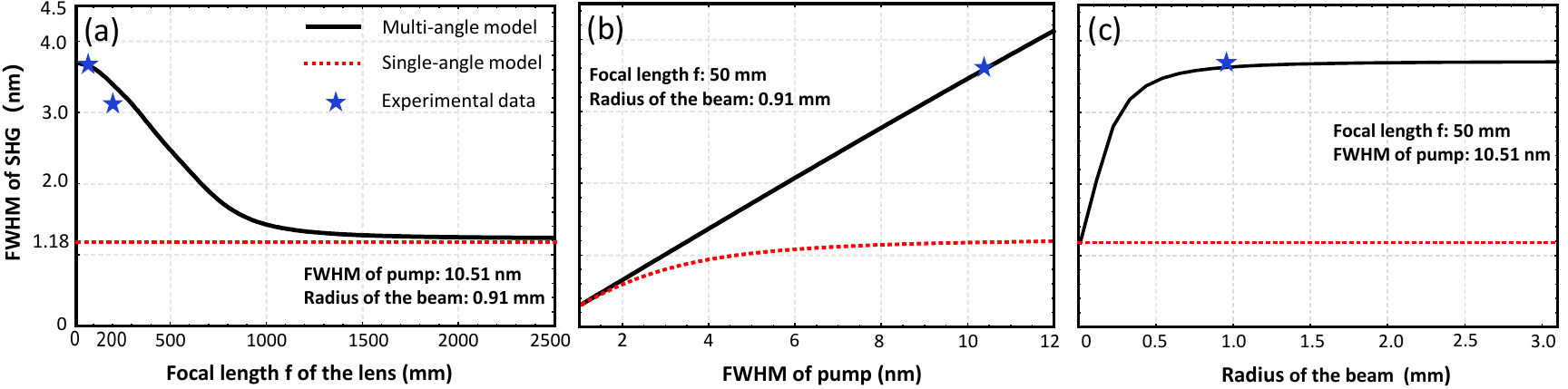}
\caption{
(a) The FWHM of the SHG laser as a function of the lens focal length (pump FWHM: 10.51 nm, pump beam radius: 0.91 mm).
(b) The FWHM of the SHG laser as a function of the pump laser FWHM (focal length: 50 mm, pump beam radius: 0.91 mm).
(c) The FWHM of the SHG laser as a function of the pump beam radius (focal length: 50 mm, pump FWHM: 10.51 nm).
In each figure, the star denotes the experimental result; \rev{ note that the error bar is smaller than the marker.} The solid black curve corresponds to the theoretical model considering multiple incident angles, while the dashed red curve corresponds to the theoretical model considering only the single incident angle $\theta_0$. \rev{The legend in panel (a) is also applicable to panels (b) and (c).}}
\label{Fig::simulation}
\end{figure*}

We  calculate the spectral distribution of the SHG using the formula below: 
\begin{equation}
g(2\omega)_{\text{total}} = \sum_{n=1}^{N}\gamma_n g_n(2\omega), 
\end{equation}
where $\gamma_n$ represents the amplitude  distribution, and
\begin{equation}
\gamma_n = \exp\left({-\frac{y_n^2}{\text{w}^2}}\right),
\end{equation}
with  \text{w} as the beam radius and 
 $y_n$  denoting the $n^{th}$ position on the y axis. The corresponding intensity distribution is
\begin{equation}
\left|\gamma_n  \right|^2 = \exp\left[{-\frac{y_n^2}{(\text{w}/\sqrt{2})^2}}\right].
\end{equation}
The final spectral intensity is 
\begin{equation}
\left|g(2\omega)_{\text{total}} \right|^2= 
\left[ \sum_{n=1}^{N}\gamma_n g_n(2\omega)\right]^2.
\end{equation}
Specifically, the spectral distribution can be calculated by first selecting the  initial parameters, such as the cutting angle $\theta_0$ of the BBO crystal, the beam radius $w$ (for example, $w = 0.91$~mm), and the lens focal length (e.g., $f = 50$~mm). The beam cross-section is then divided into $n$ strips along the $y$ axis (e.g., $n = 41$), allowing for a more detailed analysis of spatial variation. For each strip, the incident angle $\theta_n$ is determined using Eq.~(7), followed by the calculation of $g_n(2\omega)$ using Eq.~(2) and the amplitude $\gamma_n$ from Eq.~(10). By summing the contributions from all strips according to Eq.~(12), the overall spectral distribution is obtained.

We  investigate the dependence of the SHG FWHM on the focal length $f$, as shown in Fig.~3(a). The FWHM decreases monotonically with increasing focal length. This behavior can be intuitively understood: as the focal length increases, the incident angle   $\theta_2$   decreases, thereby reducing spectral broadening. we examine the effect of the pump bandwidth in Fig.~3(b), which shows an almost linear relationship between the SHG FWHM and the pump bandwidth.
 Fig.~3(c) illustrates the influence of the pump beam radius on the SHG FWHM. It is apparent that the FWHM increases monotonically as the pump beam radius grows.

We also compare the theoretical model considering only a single incident angle, $\theta_0$, represented by the red dashed lines in Fig.~3(a–c). The red dashed lines remain constant and coincide with the multi-angle model when the focal length is sufficiently large in (a), the pump bandwidth is sufficiently small in (b), or the beam radius approaches zero in (c). However, the single-incident-angle model deviates significantly from the multi-angle model under other conditions,  highlighting the necessity of using the multi-angle model for accurately describing the experiment.

\section{Experiment and Results}

\begin{figure*}[!thp]
\centering
\includegraphics[width= 1\textwidth]{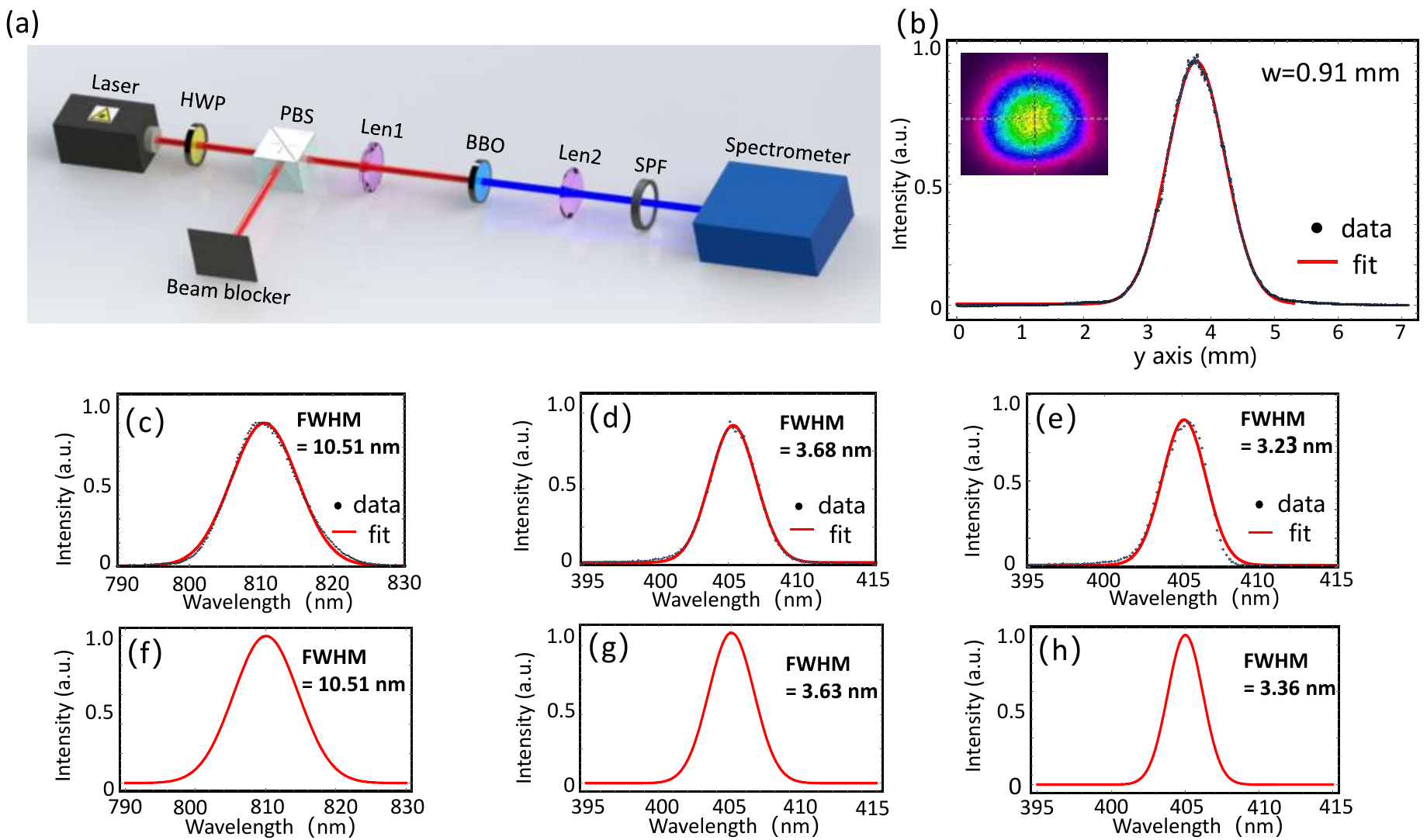}
\caption{
(a) Experimental setup.  HWP = half-wave plate; PBS = polarization beam splitter; BBO = beta barium borate; SPF = short-pass filter.
(b) Laser beam cross-section measured with a beam profiler (upper-left inset), and its vertical projection (main panel), showing a beam radius of 0.91 mm.
(c) Measured spectrum of the pump laser.
(d) Measured SHG spectrum using a lens with a focal length of 50 mm.
(e) Measured SHG spectrum using a lens with a focal length of 200 mm.
(f) Simulated spectrum of the pump laser.
(g) Calculated SHG spectrum based on Eq.~(12) for a focal length of 50 mm.
(h) Calculated SHG spectrum for a focal length of 200 mm.
}
\label{fig::result}
\end{figure*}

To validate the theoretical model, an SHG experiment was conducted using the setup illustrated in Fig.~4(a). A femtosecond Ti:sapphire laser (Mira-900, Coherent Co.), operating at a repetition rate of 76 MHz with a pulse width of approximately 100 fs, a central wavelength of 810 nm, and a spectral FWHM of 10.51 nm, served as the fundamental laser source. The laser power was adjusted using a combination of a half-wave plate (HWP) and a polarization beam splitter (PBS). After collimation, the laser beam was focused onto a BBO crystal (Type I, cut angle 28.9°, thickness 2.0 mm) by Lens-1. The generated SHG beam was subsequently collimated by Lens-2, passed through a short-pass filter (SPF), and then directed to a spectrometer (BIM-6001, Brolight Co.) for measurement.
The size of the pump beam at the position of the first lens (Len1) was measured using a laser beam profiler (LBP2, Newport Co.). Figure~\ref{fig::result}(b) shows the beam cross-section and its projection onto the vertical axis. Fitting the data with the model described in Eq.~(11) yielded a beam radius of w=0.91 mm.

The spectrum of the pump laser was measured, revealing a full width at half maximum (FWHM) of 10.51 nm, as shown in Fig.~\ref{fig::result}(c). The spectrum of the SHG output was then measured for two different lens focal lengths. Figures~\ref{fig::result}(d) and (e) display the typical spectra obtained with focal lengths of 50 mm and 200 mm, respectively, with measured FWHM values of 
3.68 nm and 3.23 nm. Statistical analysis over five independent measurements yielded mean FWHM values of 3.63$\pm$0.09 nm and 3.17$\pm$0.08 nm,respectively.   For theoretical validation, the SHG spectra were also calculated using Fig.~\ref{fig::result}(f) and Eq.~(12) for both focal lengths, yielding calculated FWHM values of 3.63 nm and 3.36 nm, as shown in Fig.~\ref{fig::result}(g) and (h), respectively. 
The close agreement between the theoretical and experimental results confirms the validity of our model.

For comparison with the theoretical models, the experimental results are presented as blue stars in Fig.~\ref{Fig::simulation}(a-b). 
 It is clear that the multi-angle model provides a much better fit to the experimental data than the single-angle model.

\section{Discussion and conclusions}
\rev{In this work, we have demonstrated both theoretically and experimentally that spatially focusing the pump beam provides an effective means to shape the spectral width of second harmonic generation (SHG). Our theoretical model describes the dependence of SHG spectral width on lens focal length, pump bandwidth, and beam radius, and was validated using BBO crystals with different focusing conditions. The results confirm that tighter focusing (shorter focal length) leads to spectral broadening, while looser focusing results in spectral narrowing. This approach offers a straightforward and practical method for precise spectral shaping, which is important for applications in microscopy, spectroscopy, and quantum optics where tailored spectral properties are required.}

\rev{Compared to previous theoretical models \cite{Bjorkholm1966PR,blit1979wavelength,boyd1968parametric} that used a Gaussian beam approximation for the laser and a rectangular crystal model, our work employs a geometric ray model for the laser and a point-like model for the crystal. Additionally, we simplify the incident angle to depend only on the y-coordinate, as illustrated in Fig. 2(b). These simplifications significantly reduce the complexity of calculations while still accurately capturing the essential physics.}

\rev{Although our experiments focused on BBO crystals, the proposed method is broadly applicable to other commonly used birefringent phase-matched crystals, such as LBO and KDP. These materials exhibit similar spectral broadening effects under varying focusing conditions due to their phase-matching properties, although differences in nonlinear coefficients and dispersion may affect the degree of broadening observed. A more detailed comparison of these inherent material properties and their impact on spectral shaping could be explored in future studies.}

\rev{Furthermore, the effect of pump focusing on spectral shaping is not limited to SHG, but extends to other nonlinear optical processes such as sum-frequency generation, difference-frequency generation, optical parametric amplification, and spontaneous parametric down-conversion (SPDC) \cite{Xu2026Modulation, Chen2025Hyperentangled, couteau2018spontaneous, Li2025Sample}. For example, in SPDC, spectral broadening can improve spectral symmetry and enhance the visibility of Hong-Ou-Mandel interference \cite{tian2022, Zhu2025,pd91-dyf9}, and spatial–spectral mapping can enable the generation of frequency-entangled qudits \cite{Yang2023Spatial}.}

\jjred{The close relationship between SHG and SPDC originates from their common $\chi^{(2)}$ nonlinear interaction. Both processes are governed by the pump 
envelope function and the phase-matching function, where birefringence plays an important role in determining the phase-matching condition. For example, a birefringently phase-matched BBO crystal can support both SHG and SPDC processes under appropriate wavelength, polarization, and phase-matching configurations.}

\rev{Beyond bandwidth control, this approach allows for precise tuning of the central wavelength and spectral shape of the SHG signal, offering additional flexibility for advanced photonic applications. In nonlinear optical microscopy, broadened spectral width corresponds to narrower temporal width, enhancing temporal resolution and enabling multichannel imaging for complex samples \cite{parodi2020nonlinear,yue2011multimodal}. In quantum optics, engineering the spectral properties of SHG is crucial for generating entangled photon pairs with desired bandwidths, which can improve the performance of quantum communication and information processing systems.}

\section*{Acknowledgments}
We thank Prof. Zhi-Yuan Zhou, Prof. Shilong Liu and Mr. Yu-Hang Luo for helpful discussions. 
This work was supported by the National Natural Science Foundations of China (Grant Numbers  12574389 and 92365106), \jjred{ and the Science and Technology Program of Hunan Province (Grant No. 2026QK3015)}.


\renewcommand\thefigure{A\arabic{figure}}
\setcounter{figure}{0}

\setcounter{equation}{0}
\renewcommand\theequation{A\arabic{equation}}

\jjred{

\appendix

\section{Comparison with the previous models}

Here we would like to compare our model with the previous models in Refs.  \cite{Bjorkholm1966PR,boyd1968parametric,blit1979wavelength}.
In Ref. \cite{Bjorkholm1966PR}, Bjorkholm proposed a focused Gaussian laser beam model for SHG, and experimentally verified the model  by measuring the power of SHG with different beam waist in 1966.
Later, in 1968  Boyd and Kleinman presented a comprehensive theory on the optimization of SHG and parametric generation \cite{boyd1968parametric}.
Based on these  theoretical models,  Blit et al measured the wavelength, temperature,  and angle bandwidths in SHG of focused beams in Ref. \cite{blit1979wavelength} in 1979.
The models used in these references are fundamentally different from ours. Specifically, in terms of the pump bandwidth, the pump laser used in Refs.  \cite{Bjorkholm1966PR,boyd1968parametric,blit1979wavelength} are narrowband, while we use a broadband pump source. For the pump propagation model, Refs.  \cite{Bjorkholm1966PR,boyd1968parametric,blit1979wavelength} adopt a Gaussian beam model, whereas we use a ray-optics (geometrical optics) model. Regarding the crystal model, Refs.~\cite{Bjorkholm1966PR,boyd1968parametric,blit1979wavelength}
consider the finite crystal interaction with a bulk crystal model, whereas our model adopts a point-like approximation to simplify the calculation. Finally, for the second-harmonic bandwidth, Refs. \cite{Bjorkholm1966PR,boyd1968parametric,blit1979wavelength} generate narrowband second-harmonic light, while our scheme produces broadband second-harmonic output. These differences are summarized in Table~\ref{tab:comparison}.
\jjred{
\begin{table}[H]
    \centering
    \caption{Comparison of the models between Refs.~\cite{Bjorkholm1966PR,boyd1968parametric,blit1979wavelength} and our work.}
    \label{tab:comparison}
    \renewcommand{\arraystretch}{1.15}

    \resizebox{\columnwidth}{!}{
        \begin{tabular}{lcc}
            \hline
            & Refs.~\cite{Bjorkholm1966PR,blit1979wavelength,boyd1968parametric} & Our work \\
            \hline
            Pump spectral width
            & Narrowband
            & Broadband \\
            
            Laser propagation model
            & Gaussian beam
            & Geometrical ray\\
            
            Crystal model
            & Rectangle
            & Point-like\\
            
            SHG spectral width
            & Narrowband
            & Broadband \\
            \hline
        \end{tabular}
    }
\end{table}

}

\section{Comparison with the Boyd-Kleinman function}

The Boyd-Kleinman function is a well-known function which describes the power of SHG for the focused Gaussian beam model \cite{Bjorkholm1966PR,boyd1968parametric,blit1979wavelength,boyd2020nonlinear}. 
For comparison, the geometrical ray model developed in this work can also be formulated in a similar manner. 
In this section, we derive the corresponding function in detail. 
For a single incident angle $\theta$,  the SHG field is
\begin{equation}
E(\delta\theta)
\propto
\int_{0}^{L}
e^{i\Delta k(\delta\theta)z}\,dz
=
L\,\operatorname{sinc}
\left(
\frac{\beta L\delta\theta}{2}
\right)
e^{i\beta L\delta\theta/2},
\end{equation}
where $\delta \theta$ is the difference between the incident angle $\theta$ and the cutting angle $\theta_0$, $L$ is the length of the crystal, $z$ is the poisiton on the crystal, $\Delta k$ phase mismatch and  
\begin{equation}
\Delta k(\delta\theta)
\approx
\beta\,\delta\theta,
\end{equation}
where 
$\beta=
\left.
\frac{\partial \Delta k}{\partial \theta}
\right|_{\theta_0}$.
After integration of $E(\delta\theta)$ on all the incident angles, we obtain the total optical field G: 
\begin{equation}
G
\propto
L
\int_{-\infty}^{+\infty}
e^{-\delta\theta^{2}/\Delta\theta^{2}}
\operatorname{sinc}
\left(
\frac{\beta L\delta\theta}{2}
\right)
e^{i\beta L\delta\theta/2}
\,d\delta\theta,
\end{equation}
where 
$e^{-\delta\theta^{2}/\Delta\theta^{2}}$ is the weight coefficient.

The above equation can be simplified as 
\begin{equation}
G
\propto
\frac{L\Delta\theta\sqrt{\pi}}{2}
\int_{-1}^{1}
\exp\left[
-\frac{\Delta\theta^{2}\beta^{2}L^{2}}{16}(1+t)^{2}
\right]
\,dt.
\end{equation}
In the above calculation, we utilized the following two equations:
\begin{equation}
\operatorname{sinc}(X)e^{iX}
=
\frac{1}{2}
\int_{-1}^{1}
e^{iX(1+t)}\,dt.
\end{equation}

\begin{equation}
\int_{-\infty}^{+\infty}
e^{-ax^{2}+ibx}\,dx
=
\sqrt{\frac{\pi}{a}}
\exp\left(
-\frac{b^{2}}{4a}
\right).
\end{equation}

By introducing the  error function $\operatorname{erf}$, $G$ can be further simplified as 
\begin{equation}
G
\propto
\frac{\pi}{\beta}
\operatorname{erf}
\left(
\frac{\Delta\theta\beta L}{2}
\right).
\end{equation}
The power of SHG $P_{2\omega}=K|G|^2$ can be obtained as
\begin{equation}
P_{2\omega}
=
K\frac{\pi^{2}}{\beta^{2}}
\operatorname{erf}^{2}
\left(
\frac{\Delta\theta\beta L}{2}
\right),
\end{equation}
where $K$ is a constant determined by the pump power, effective nonlinear coefficient, etc. 
By  introducing the parameter  
\begin{equation}
\xi
=
\frac{\Delta\theta\beta L}{2},
\end{equation}
and 
\begin{equation}
h
=
\frac{\pi^{2}\Delta\theta}{2\beta\xi}
\operatorname{erf}^{2}(\xi), 
\end{equation} we can obtain
\begin{equation}
P_{2\omega}=K\times L\times h,
\end{equation}
where $h$ plays a role similar to the Boyd--Kleinman function in describing
the focusing dependence of SHG efficiency for the focused Gaussian beam model
\cite{boyd1968parametric,boyd2020nonlinear}.

\jjred{

\section{Comparison under highly tight focusing conditions}

In this section, we further investigate the case of highly non-paraxial 
beams, corresponding to tight focusing conditions, where the focal length 
of the lens is shorter than 50~mm.

\begin{figure*}[!htbp]
\centering
\includegraphics[width=1\linewidth]{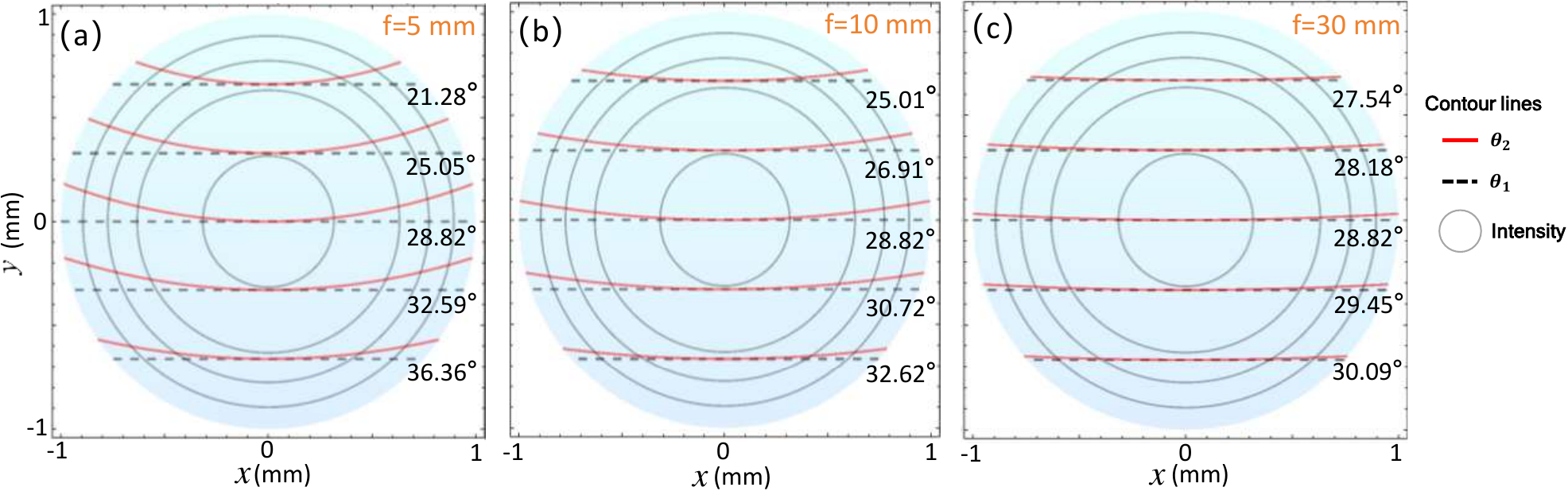}
\caption{
Calculated distributions of the incident angles $\theta_1$ and $\theta_2$
across the cross section of the pump beam under different focal lengths:
(a) $f=5$~mm, (b) $f=10$~mm, and (c) $f=30$~mm.
}
\label{fig:A1}
\end{figure*}

Figure~\ref{fig:A1} shows the calculated distribution of the incident angles 
$\theta_1$ and $\theta_2$ across the cross section of the pump beam for 
 focal lengths of 5~mm, 10~mm, and 30~mm. It can be observed that 
the deviation between $\theta_1$ and $\theta_2$ gradually increases as the 
focal length decreases, indicating the increasing influence of non-paraxial 
effects under tighter focusing conditions.

\begin{table}[!thp]
\centering
\renewcommand{\arraystretch}{1.2}
\setlength{\tabcolsep}{4pt}
\begin{tabular}{cccccc}
\hline
$f$ (mm) 
& 5 & 10 & 20 & 30 & 40 \\
\hline
FWHM ($\theta_1$) (nm)
&3.901&3.714&3.710&3.705&3.697\\
FWHM ($\theta_2$) (nm)
&3.727&3.714&3.710&3.705&3.697\\
\hline
\end{tabular}
\caption{
Calculated FWHM of the SHG spectra based on $\theta_1$ and $\theta_2$
for different focal lengths.
}
\label{tab:A2}
\end{table}

Table~\ref{tab:A2} compares the calculated FWHM of the SHG spectra obtained
using $\theta_1$ (Eq.~(7)) and $\theta_2$ (Eq.~(8)). The calculation
parameters are a pump bandwidth of 10.51~nm, a crystal length of 2~mm,
and a beam radius of w=0.9~mm,
for a BBO crystal.

The results show that the calculated FWHM values are almost identical for
focal lengths ranging from 10~mm to 40~mm. However, a noticeable difference
appears when the focal length is reduced to 5~mm. This difference originates
from the Gaussian intensity distribution of the pump beam. Although the
angular deviation becomes significant near the beam edge under tight focusing
conditions, the corresponding optical intensity is relatively low, and
therefore these regions contribute less to the SHG process.

In summary, the results demonstrate that our model remains valid even under
highly tight focusing conditions. Although non-paraxial effects become more
significant for extremely short focal lengths, their contribution to the
generated SHG spectrum remains limited due to the Gaussian intensity weighting
and the finite effective angular distribution determined by the numerical aperture.

\section{Relation to the numerical aperture (NA)}

The tight focusing condition is closely related to the numerical aperture (NA),
which determines the angular distribution of the focused pump beam. For a focusing lens with focal length $f$ and an effective beam radius w, the effective numerical aperture in air can be described as

\setcounter{equation}{11}

\begin{equation}
NA_{\mathrm{eff}}
=
\sin\left[\arctan\left(\frac{\text{w}}{f}\right)\right]
=
\frac{\text{w}}{\sqrt{f^2+\text{w}^2}} .
\end{equation}

Under the paraxial approximation (w$\ll f$), this expression can be simplified
as $NA_{\mathrm{eff}}\approx$ w$/f$. A larger value of w$/f$ corresponds to a
larger numerical aperture and a broader angular distribution of the pump beam.
Since different incident angles experience different phase-matching
conditions, the numerical aperture directly affects the angle-dependent
spectral response of SHG.

}
}


\bibliography{2023LatexTemplate}
\end{document}